\documentclass[lettersize,journal]{IEEEtran}
\usepackage{amsmath,amsfonts}
\usepackage{algorithmic}
\usepackage{algorithm}
\usepackage{array}
\usepackage[caption=false,font=normalsize,labelfont=sf,textfont=sf]{subfig}
\usepackage{textcomp}
\usepackage{stfloats}
\usepackage{url}
\usepackage{verbatim}
\usepackage{graphicx}
\usepackage{cite}
\usepackage{hyperref}
\usepackage{subcaption}
\usepackage[left=0.5in,right=0.5in,top=0.5in,bottom=0.5in]{geometry}
\usepackage[font=small,labelfont=small]{caption}

\begin{document}

\title{Joint Antenna Placement and Power Allocation for RSMA-Enabled Pinching Antenna Systems}

\author{
Harsh~Raj,
Mallena~Vardhan, Keshav~Singh,~\textit{Senior Member,~IEEE}, and Sudip~Biswas,~\textit{Member,~IEEE}%
\thanks{Harsh Raj, Mallena Vardhan and Sudip Biswas are with the Department of Electronics and Communication Engineering, Indian Institute of Information Technology Guwahati, Guwahati, Assam 781015, India (e-mail: \{harsh.raj23b, mallena.vardhan23b\}@iiitg.ac.in; sudip.biswas@ieee.org).}%
\thanks{Keshav Singh is with the Institute of Communications Engineering, National Sun Yat-sen University, Kaohsiung 80424, Taiwan (e-mail: keshav.singh@mail.nsysu.edu.tw).}%

}

\maketitle

\begin{abstract}
This letter investigates a rate-splitting multiple access (RSMA)-enabled multi-user pinching antenna system (PASS). A fairness-aware sum-rate maximization problem is formulated to jointly optimize pinching antenna locations and common/private stream power allocation. The resulting mixed discrete-continuous non-convex problem is addressed using an alternating optimization framework that combines greedy antenna placement with successive convex approximation (SCA)-based power allocation. Numerical results demonstrate that the proposed RSMA-enabled PASS significantly improves achievable sum-rate, user fairness, and bit error rate (BER) performance compared with conventional non-RSMA PASS schemes.
\end{abstract}
\vspace{-0.25em}

\vspace{-1em}
\section{Introduction}
\IEEEPARstart{P}{inching} antenna systems (PASS) have recently emerged as a promising electromagnetic-domain transmission architecture for future wireless networks, offering a fundamentally different design paradigm from conventional antenna arrays \cite{11263923}. Unlike traditional multi-antenna systems that require dedicated radio-frequency (RF) chains and complex feeding networks, PASS utilizes a shared dielectric waveguide along which multiple pinching antennas can be dynamically positioned to extract and radiate electromagnetic energy \cite{Liu2025PASS_Mag,Liu2025PASS_Principles}. By enabling flexible spatial deployment of radiating elements over a common waveguide, PASS introduces a new degree of freedom in wireless system design, allowing the propagation environment to be directly shaped through antenna positioning. This unique capability has demonstrated significant potential for improving spectral efficiency, energy efficiency, and hardware scalability in future high-frequency communication systems. Moreover, the ability to flexibly position radiating elements makes PASS particularly attractive for near-field (NF) communications, where highly localized beam focusing and spatially selective signal delivery can be exploited to enhance user performance \cite{Chen2025PASS_Rate}.

The performance gains of PASS, however, critically depend on the spatial configuration of pinching antennas. Since multiple users are simultaneously served through a limited number of radiating elements coupled to the same waveguide, the resulting channels are inherently correlated and interference-limited. Existing PASS studies have primarily focused on channel modeling, waveguide design, antenna placement strategies, and beamforming optimization. Nevertheless, the problem of efficiently managing multi-user interference in PASS remains largely unexplored. As user density increases, conventional linear transmission schemes become increasingly susceptible to residual interference, limiting the achievable multiplexing gains offered by the spatial flexibility of PASS.

Rate-splitting multiple access (RSMA) is a powerful interference management framework that bridges the performance gap between fully decoding interference and fully treating interference as noise \cite{9936974,Clerckx2023Primer}. By splitting user messages into common and private streams, RSMA enables partial interference decoding through successive interference cancellation (SIC), thereby improving robustness against channel coupling, imperfect precoding, and heterogeneous user conditions \cite{8744489,Mao2018RSMA}. These properties make RSMA particularly attractive for PASS architectures, where the spatially reconfigurable antenna locations directly influence channel correlations and inter-user interference patterns. However, despite the rapid development of both PASS and RSMA technologies, their joint integration remains largely unexplored in the literature.

Motivated by the complementary benefits of PASS and RSMA, this letter investigates an RSMA-enabled multi-user PASS downlink system, where multiple pinching antennas are deployed along a single dielectric waveguide to serve multiple users simultaneously. By jointly exploiting the spatial reconfigurability of pinching antennas and the interference management capability of RSMA, the proposed framework enhances spectral efficiency and user fairness in interference-limited environments. The main contributions of this letter are summarized as follows:
\begin{itemize}

\item We propose a RSMA-enabled PASS architecture for multi-user downlink communications, where the spatial placement of pinching antennas and the common/private message transmission strategy are jointly incorporated into the system design.

\item We formulate a fairness-aware sum-rate maximization problem that jointly optimizes the pinching antenna locations and the power allocation of common and private streams under transmit power, quality-of-service (QoS), and practical antenna placement constraints.

\item The resulting mixed discrete-continuous non-convex optimization problem is addressed through an alternating optimization successive convex approximation (AO-SCA) framework. Specifically, a low-complexity grid-based greedy strategy is employed for pinching antenna placement, while the power allocation subproblem is reformulated as a  difference-of-conve (DC) program and solved via SCA.

\item Simulation results demonstrate that the proposed RSMA-enabled PASS framework achieves superior sum-rate, fairness, and bit error rate (BER) performance compared with conventional non-RSMA PASS schemes, highlighting the benefits of jointly leveraging spatial reconfigurability and advanced interference management.
\end{itemize}

\section{System Model}

\subsection{Pinching antenna setup}

\begin{figure}[t]
    \centering
    \includegraphics[width=\linewidth]{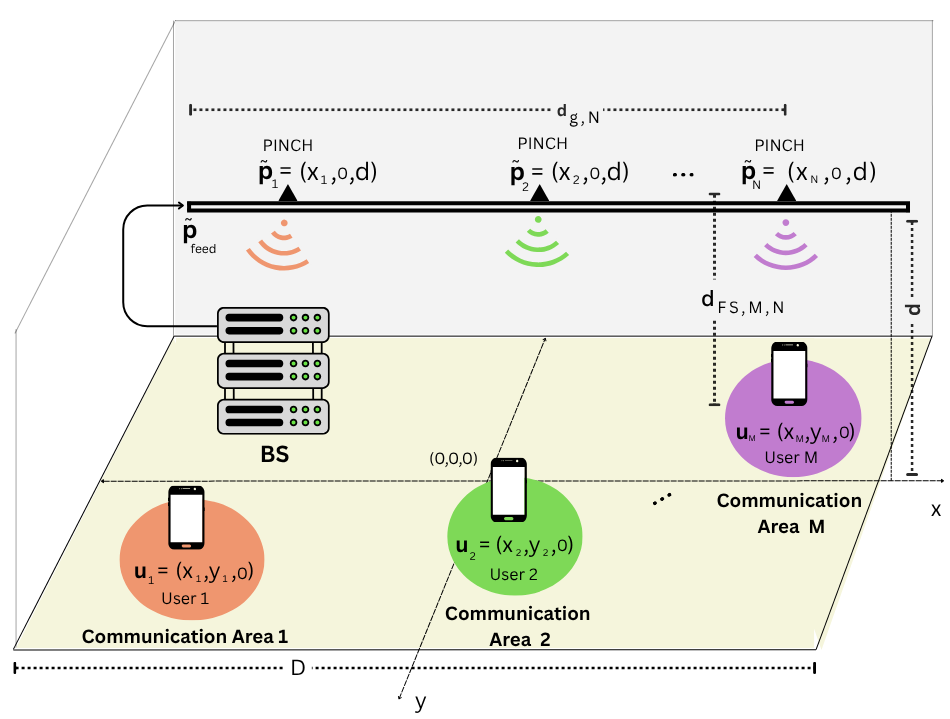}
    \caption{Illustration of  the system model of the proposed scheme.}
    \label{fig:system_model}
    \vspace{-1.25em}
\end{figure}

We consider a downlink PASS, illustrated in Fig.~\ref{fig:system_model}, where a base station (BS) equipped with $N$ pinching antennas, denoted by $\mathcal{N}=\{1,2,\ldots,N\}$, serves $M$ single-antenna users, denoted by $\mathcal{M}=\{1,2,\ldots,M\}$. The antennas radiate the signal guided along the dielectric waveguide to establish line-of-sight (LoS) links with the users.
The system operates in a three-dimensional Cartesian coordinate system where the waveguide is aligned along the $x$-axis at height $d$. Users are randomly distributed in a square service area of side length $D$ on the $x$–$y$ plane. The location of the $m$-th user is denoted by ${{u}_m}=(x_m,y_m,0)$ with $x_m,y_m\in[-\frac{D}{2},\frac{D}{2}]$. The feeding point of the waveguide is $\tilde{p}_{\text{feed}}=(-\frac{D}{2},0,d)$, while the position of the $n$-th pinching antenna is ${\tilde{p}}_n=(x_n,0,d)$.
The free-space distance between user $m$ and antenna $n$, and the propagation distance along the waveguide, are given by
$
d_{F,m,n}=\|{\tilde{p}}_n-{u}_m\|,\quad
d_{g,n}=\|{\tilde{p}}_n-\tilde{p}_{\text{feed}}\|.
$
Accordingly, the channel vector $\mathbf{h}_m\in\mathbb{C}^{N\times1}$ is
\begin{equation}
\mathbf{h}_m=
\sqrt{\eta}
\begin{bmatrix}
\dfrac{e^{-j2\pi\left(\frac{d_{F,m,1}}{\lambda}+\frac{d_{g,1}}{\lambda_g}\right)}}{d_{F,m,1}},
\cdots,
\dfrac{e^{-j2\pi\left(\frac{d_{F,m,N}}{\lambda}+\frac{d_{g,N}}{\lambda_g}\right)}}{d_{F,m,N}}
\end{bmatrix},
\end{equation}
where $\eta$ is the coupling efficiency, $\lambda$ is the carrier wavelength, and the waveguide attenuation is neglected.
Since all antennas radiate the same guided signal, the effective scalar channel for user $m$ is obtained by coherently combining the contributions from all pinching antennas. The factor $\sqrt{1/N}$ normalizes the transmit power and assumes equal signal leakage from each pinching antenna \cite{10945421}.
\begin{equation}
h_m^{\text{eff}}=
\sqrt{\frac{1}{N}}
\sum_{n=1}^{N}
\frac{\sqrt{\eta}}{d_{F,m,n}}
e^{-j2\pi\left(\frac{d_{F,m,n}}{\lambda}+\frac{d_{g,n}}{\lambda_g}\right)}.
\end{equation}

\subsection{Signal model}
The transmit signal $x$ injected at the feeding point ${\tilde{p}}_{\text{feed}}$ follows the single-layer RSMA principle. It consists of a superposition of one common stream $s_c$ and $M$ private streams $s_m$:
\begin{equation}
x = \sqrt{P_c}\, s_c + \sum_{m=1}^{M} \sqrt{P_m}\, s_m,
\end{equation}
where $s_c$ and $s_m$ are independent and identically distributed (i.i.d.) complex Gaussian symbols with unit average power, i.e., $\mathbb{E}[|s_c|^2] = \mathbb{E}[|s_m|^2] = 1\quad \forall m \in \mathcal{M}$. The transmit power allocation satisfies
$
P_c + \sum_{m=1}^{M} P_m \le P_{\text{total}},
$
where $P_c$ denotes the power allocated to the common stream and $P_m$ denotes the power allocated to the private stream intended for user $m$.
The received signal at user $m$ is given by
\begin{equation}
y_m = h_m^{\text{eff}} x + n_m,
\end{equation}
where $h_m^{\text{eff}}$ represents the effective channel between the transmitter and user $m$, and $n_m \sim \mathcal{CN}(0,\sigma^2)$ denotes additive white Gaussian noise (AWGN) with variance $\sigma^2$.

\vspace{-1em}

\subsection{Decoding protocol and achievable rates}
The RSMA scheme follows a single-layer decoding structure, where the common message is decoded by all users \cite{Yang2020RSMA_SumRate}.

\subsubsection{Common stream decoding}

All users first decode the common stream $s_c$ while treating all private streams as noise. The signal-to-interference-plus-noise ratio (SINR) for decoding the common stream at user $m$ is given by
\begin{equation}
\gamma_{c,m} =
\frac{P_c |h_m^{\text{eff}}|^2}
{|h_m^{\text{eff}}|^2 \sum_{j=1}^{M} P_j + \sigma^2}
\label{eqn:sinr_common},
\end{equation}
where the denominator accounts for interference from all private streams and noise power. To ensure successful decoding by all users, the common rate is limited by the worst-user channel as
$
R_c = B \min_{m \in \mathcal{M}}
\left\{ \log_2(1 + \gamma_{c,m}) \right\}.
$
The common rate is divided among users such that
$
R_c = \sum_{m=1}^{M} C_m,
$
where $C_m$ denotes the portion of the common rate allocated to user $m$.

\subsubsection{Private stream decoding}

After decoding $s_c$, each user applies SIC to remove the common signal from $y_m$. User $m$ then decodes its private stream $s_m$ while treating the remaining private streams as interference. The SINR for decoding $s_m$ at user $m$ is given by
\begin{equation}
\gamma_{p,m} =
\frac{P_m |h_m^{\text{eff}}|^2}
{|h_m^{\text{eff}}|^2 \sum_{j \ne m} P_j + \sigma^2}
\label{eqn:sinr_private}.
\end{equation}
The achievable private rate for user $m$ is given by
$R_{p,m} = B \log_2(1 + \gamma_{p,m}).
$
The total achievable rate for user $m$ is therefore expressed as
$R_m = C_m + R_{p,m}.$
The overall system sum-rate is given by
$R_{\text{sum}} = \sum_{m=1}^{M} R_m = \sum_{m=1}^{M} (C_m + R_{p,m}).$

\section{Problem Formulation}
In this section, we formulate the joint optimization problem for the proposed RSMA-enabled PASS architecture. The objective is to jointly optimize the transmit power allocation and pinching antenna positions to maximize the achievable system sum rate while satisfying communication and physical constraints.
Let
$
\mathbf{P}=[P_c,P_1,P_2,\ldots,P_M]^T
$
denote the transmit power allocation vector, where $P_c$ and $P_m$ represent the powers allocated to the common stream and the private stream of user $m$, respectively. Furthermore, let
$
\tilde{\mathbf{x}}=[x_1,x_2,\ldots,x_N]^T
$
denote the pinching antenna position vector, where $x_n$ represents the location of the $n$-th pinching antenna along the waveguide of length $D$.

For a given $(\mathbf{P},\tilde{\mathbf{x}})$, the achievable common-stream and private-stream rates of user $m$ are denoted by $R_c(\mathbf{P},\tilde{\mathbf{x}})$ and $R_{p,m}(\mathbf{P},\tilde{\mathbf{x}})$, respectively. The objective is to jointly determine $\mathbf{P}$ and $\tilde{\mathbf{x}}$ to maximize the overall system throughput.
Accordingly, the joint optimization problem is formulated as 
\begin{subequations}\label{eq:JointProblem}
\begin{align}
\max_{\mathbf{P},\tilde{\mathbf{x}}} \quad 
& R_c(\mathbf{P},\tilde{\mathbf{x}}) + \sum_{m\in\mathcal{M}} R_{p,m}(\mathbf{P},\tilde{\mathbf{x}}) \\
\text{s.t.} \quad 
& P_c + \sum_{m\in\mathcal{M}} P_m \le P_{\text{total}}, \label{cons:PowerBudget} \\
& R_c(\mathbf{P},\tilde{\mathbf{x}}) \ge R_{c,\min}, \label{cons:CommonRate} \\
& R_{p,m}(\mathbf{P},\tilde{\mathbf{x}}) \ge R_{p,\min},  \quad \forall m \in \mathcal{M}, \label{cons:PrivateRate} \\
& P_c, P_m \ge 0, \quad \forall m \in \mathcal{M}, \label{cons:NonNegativePower} \\
& x_n \in \left[-\frac{D}{2}, \frac{D}{2}\right], \quad \forall n \in \mathcal{N},\label{cons:PositionRange} \\
& |x_i - x_j| \ge \frac{\lambda}{2}, \quad \forall i,j \in \mathcal{N}, \; i \neq j.\label{cons:AntennaSpacing}
\end{align}
\end{subequations}
Constraint \eqref{cons:PowerBudget} limits the total transmit power according to the hardware capability, while \eqref{cons:CommonRate} and \eqref{cons:PrivateRate} enforce the QoS requirements for the common and private streams. The non-negativity condition in \eqref{cons:NonNegativePower} ensures feasible power allocation. Furthermore, \eqref{cons:PositionRange} restricts the pinching antennas within the waveguide of length $D$, whereas \eqref{cons:AntennaSpacing} guarantees a minimum spacing of $\lambda/2$ to mitigate strong electromagnetic coupling.

Problem \eqref{eq:JointProblem} is highly non-convex due to the coupled dependence of the achievable rates on the power allocation $\mathbf{P}$ and antenna positions $\tilde{\mathbf{x}}$. To facilitate tractable optimization, we decompose the original problem into antenna placement and power allocation subproblems, which are solved iteratively using an AO framework.

\subsubsection{Subproblem 1- (Pinching antenna position optimization)}
For a fixed power allocation $\mathbf{P}$, the optimization reduces to determining the optimal positions of the pinching antennas. The corresponding position optimization problem is formulated as
\begin{subequations}\label{eq:PositionSubproblem}
\begin{align}
\max_{\tilde{\mathbf{x}}} \quad 
& R_c(\tilde{\mathbf{x}}) + \sum_{m\in\mathcal{M}} R_{p,m}(\tilde{\mathbf{x}}) \\
\text{s.t.} \quad 
& \eqref{cons:PositionRange},\ \eqref{cons:AntennaSpacing}.
\end{align}
\end{subequations}
The achievable rates depend on the antenna positions through the channel coefficients, making the objective function nonlinear with respect to $\tilde{\mathbf{x}}$. In addition, the spacing constraint enforces a minimum separation of $\lambda/2$ between antennas, resulting in a non-convex position optimization problem.

\subsubsection{Subproblem 2- (Power allocation optimization)}
For fixed antenna positions $\tilde{\mathbf{x}}$, the channel gains remain constant and the joint optimization reduces to determining the optimal transmit power allocation. The resulting power allocation subproblem is formulated as
\begin{subequations}\label{eq:PowerSubproblem}
\begin{align}
\max_{\mathbf{P}} \quad 
& R_c(\mathbf{P}) + \sum_{m\in\mathcal{M}} R_{p,m}(\mathbf{P}) \\
\text{s.t.} \quad 
 & \eqref{cons:PowerBudget},\ \eqref{cons:CommonRate},\ \eqref{cons:PrivateRate},\ \eqref{cons:NonNegativePower}.
\end{align}
\end{subequations}
Although the antenna positions are fixed, problem \eqref{eq:PowerSubproblem} remains non-convex due to the coupled SINR-dependent rate expressions. In the next section, this problem is reformulated using a DC representation and solved via SCA.

\vspace{-0.7em}

\section{Proposed Solution}

\subsection{Greedy pinching antenna placement}

To obtain feasible antenna positions with low computational complexity, a grid-based greedy placement strategy is adopted. Specifically, the waveguide is discretized into a set of candidate locations with spacing $\lambda/2$. Each user is then associated with one pinching antenna positioned at the nearest feasible grid point while satisfying the minimum spacing constraint. This placement strategy improves the effective channel gains by reducing the combined waveguide propagation and wireless transmission distances.

\subsection{SCA-based power allocation}
For fixed antenna positions $\tilde{\mathbf{x}}$, the effective channel gains $g_m = |h_m^{\text{eff}}|^2$ remain constant, and the resulting power allocation problem exhibits a DC structure. We therefore employ an SCA-based approach, where the non-convex interference terms are replaced by their first-order approximations around the current operating point.

\vspace{0.4em}
\subsubsection{DC decomposition}

The private rate of user $m$ can be expressed in a DC form as
\begin{equation}
R_{p,m}(\mathbf{P})
=
\underbrace{\log_2 \!\left( \sum_{j=1}^{M} P_j  g_m + \sigma^2 \right)}_{f_{p,m}(\mathbf{P})}
-
\underbrace{\log_2 \!\left( \sum_{j \neq m} P_j g_m + \sigma^2 \right)}_{g_{p,m}(\mathbf{P})}.
\end{equation}

\begin{equation}
R_c(\mathbf{P}) =
\underbrace{\log_2 \left( \sum_{j=0}^{M} P_j g_{\min} + \sigma^2 \right)}_{f_c(\mathbf{P})}
-
\underbrace{\log_2 \left( \sum_{j=1}^{M} P_j g_{\min} + \sigma^2 \right)}_{g_c(\mathbf{P})}.
\end{equation}
Both $f_{p,m}(\mathbf{P})$ and $g_{p,m}(\mathbf{P})$ are concave with respect to $\mathbf{P}$. The subtraction of the concave interference term $g_{p,m}(\mathbf{P})$ makes the rate expression non-concave. Similarly, the common rate can be expressed in a DC form, where 
$f_c(\mathbf{P})$ and $g_c(\mathbf{P})$ denote the signal and interference
components of the common rate expression, respectively.

\vspace{0.4em}
\subsubsection{SCA-based linearization}
Let $\mathbf{P}^{(k)}$ denote the power allocation at the $k$-th SCA iteration. The non-convex interference terms are approximated via first-order Taylor expansions around $\mathbf{P}^{(k)}$. For the common stream, the linearized interference term is
\begin{equation}
\begin{aligned}
\hat{g}_c(\mathbf{P};\mathbf{P}^{(k)})
&= g_c(\mathbf{P}^{(k)}) \\
& + \frac{g_{\min}}{\ln(2)\left(I_c^{(k)}+\sigma^2\right)}
\sum_{j=1}^{M}(P_j - P_j^{(k)}),
\end{aligned}
\end{equation}
where $I_c^{(k)} = g_{\min}\sum_{j=1}^{M} P_j^{(k)}$, and $g_{\min}$ denotes the effective channel gain of the user with the minimum common decoding capability,
i.e., $g_{\min} = \min_{m \in \mathcal{M}} g_m$.
Similarly, the linearized interference term for the private stream of user $m$ is given by
\begin{equation}
\begin{aligned}
\hat{g}_{p,m}(\mathbf{P};\mathbf{P}^{(k)})
&= g_{p,m}(\mathbf{P}^{(k)}) \\
&+ \frac{g_m}{\ln(2)\left(I_{p,m}^{(k)}+\sigma^2\right)}
\sum_{j \neq m}(P_j - P_j^{(k)}),
\end{aligned}
\end{equation}
where $I_{p,m}^{(k)} = g_m \sum_{j \neq m} P_j^{(k)}$.
Based on these approximations, the surrogate rate functions are defined as
\begin{align}
\hat{R}_c(\mathbf{P};\mathbf{P}^{(k)}) &= f_c(\mathbf{P}) - \hat{g}_c(\mathbf{P};\mathbf{P}^{(k)}), \\
\hat{R}_{p,m}(\mathbf{P};\mathbf{P}^{(k)}) &= f_{p,m}(\mathbf{P}) - \hat{g}_{p,m}(\mathbf{P};\mathbf{P}^{(k)}).
\end{align}
Since $f(\cdot)$ is concave and $\hat{g}(\cdot)$ is affine, the surrogate rate functions are concave, yielding a convex approximation of the original problem at each iteration.

\vspace{0.4em}
\subsubsection{Convex subproblem}

At iteration $k$, the power allocation is updated by solving the following convex optimization problem
\begin{subequations} \label{eqn:Sub1_CVX_Final}
\begin{align}
\max_{\mathbf{P}} \quad 
& \hat{R}_c(\mathbf{P};\mathbf{P}^{(k)}) 
+ \sum_{m \in \mathcal{M}} \hat{R}_{p,m}(\mathbf{P};\mathbf{P}^{(k)}) \\
\text{s.t.} \quad 
& \hat{R}_c(\mathbf{P};\mathbf{P}^{(k)}) \ge R_{c,\min}, \\
& \hat{R}_{p,m}(\mathbf{P};\mathbf{P}^{(k)}) \ge R_{p,\min}, \quad \forall m \in \mathcal{M}, \\
& \eqref{cons:PowerBudget},\ \eqref{cons:NonNegativePower}.
\end{align}
\end{subequations}
The resulting problem is convex and can be efficiently solved using CVX. The SCA iterations generate a non-decreasing objective sequence and converge to a stationary point.

The overall AO-SCA procedure for jointly optimizing the pinching antenna locations and transmit power allocation is summarized in Algorithm~\ref{alg:AO_SCA}.

\vspace{-0.4em}
\subsection{Convergence and complexity}

The proposed AO-SCA framework alternately updates the pinching antenna locations and transmit power allocation. For fixed antenna positions, the SCA procedure constructs a sequence of convex approximations that generate a non-decreasing objective value across iterations. Since the achievable sum rate is upper bounded by the finite transmit power budget, the objective sequence is guaranteed to converge. Consequently, the overall AO-SCA framework converges to a block-wise stationary solution.

The computational complexity comprises the greedy antenna placement step and the SCA-based power allocation step. The former incurs a complexity of $\mathcal{O}(N|\mathcal{G}|)$, where $|\mathcal{G}|$ denotes the number of candidate grid locations. The latter involves solving a convex problem with $M+1$ optimization variables via an interior-point method, resulting in a complexity of $\mathcal{O}((M+1)^{3.5})$ per SCA iteration \cite{Boyd2004Convex}. Therefore, the overall complexity of the proposed algorithm is
\begin{equation}
\mathcal{O}\!\left(
I_{\rm AO}
\left[
N|\mathcal{G}|+
I_{\rm SCA}(M+1)^{3.5}
\right]
\right),
\end{equation}
where $I_{\rm AO}$ and $I_{\rm SCA}$ denote the numbers of AO and SCA iterations, respectively.

\begin{algorithm}[t]
\caption{\small AO-SCA algorithm for joint antenna placement and power allocation}
\label{alg:AO_SCA}
\small
\begin{algorithmic}[1]
\STATE \textbf{Initialize:} Feasible power allocation $\mathbf{P}^{(0)}$, antenna positions $\tilde{\mathbf{x}}^{(0)}$, tolerance $\epsilon$, and iteration index $k=0$.
\REPEAT
    \STATE Update $\tilde{\mathbf{x}}^{(k+1)}$ using the grid-based greedy antenna placement strategy for fixed $\mathbf{P}^{(k)}$.
    
    \STATE Solve \eqref{eqn:Sub1_CVX_Final} via SCA for fixed $\tilde{\mathbf{x}}^{(k+1)}$ and obtain $\mathbf{P}^{(k+1)}$.
    
    \STATE $k \leftarrow k+1$.
\UNTIL{$|R_{\rm sum}^{(k)}-R_{\rm sum}^{(k-1)}| \le \epsilon$}
\STATE \textbf{Output:} $\tilde{\mathbf{x}}^{(k)}$ and $\mathbf{P}^{(k)}$.
\end{algorithmic}
\end{algorithm}

\begin{figure}[t]
    \centering
    \includegraphics[width=0.32\textwidth]{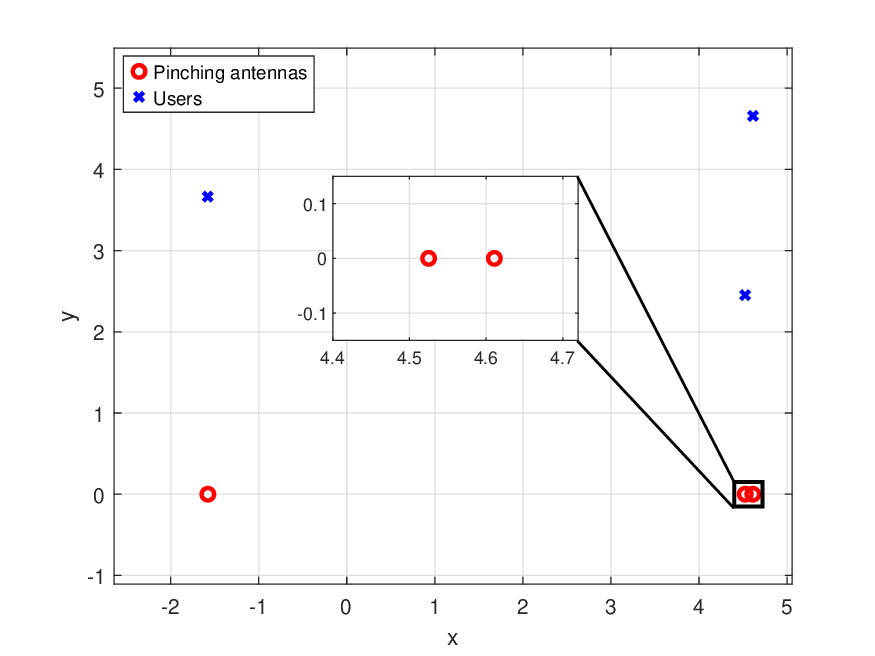}
    \caption{Optimized antenna-user geometry for $N=M=3$.}
    \label{fig:geometry}
    \vspace{-1.5em}
\end{figure}

\begin{figure*}[t]
\vspace{-1.25em}
\centering

\begin{minipage}{0.32\textwidth}
\centering
\includegraphics[width=\linewidth]{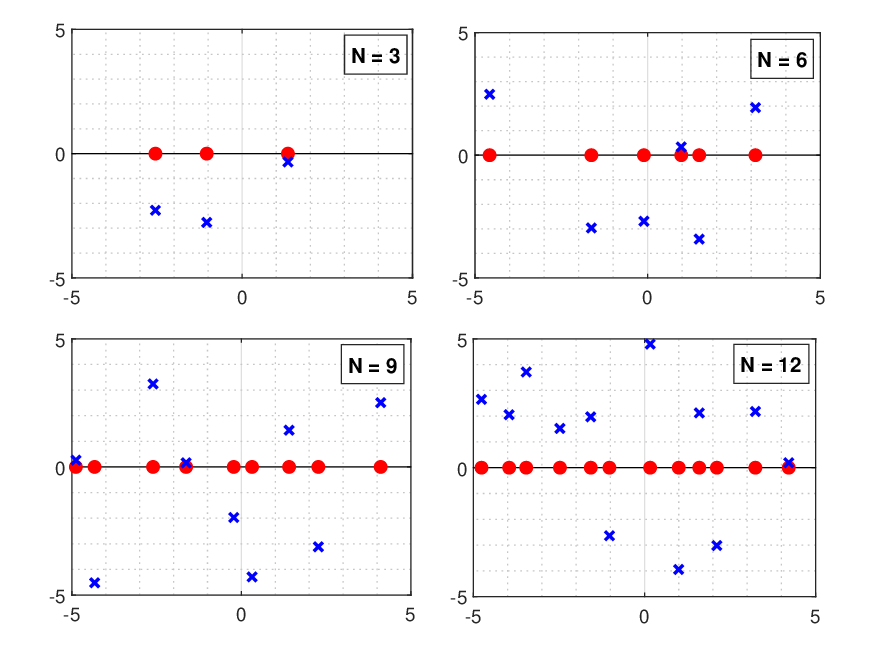}
\vspace{-0.5em}
\caption{Optimized geometries for different values of $N=M$.}
\label{fig:comparison_geo}
\end{minipage}
\hfill
\begin{minipage}{0.32\textwidth}
\centering
\includegraphics[width=\linewidth]{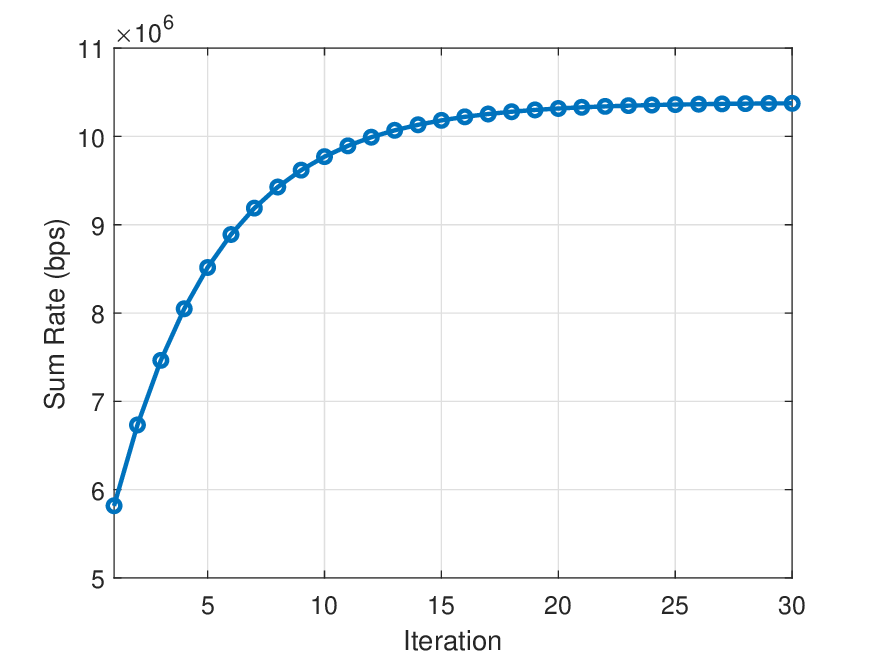}
\vspace{-0.5em}
\caption{Convergence of the proposed AO-SCA algorithm.}
\label{fig:objective}
\end{minipage}
\hfill
\begin{minipage}{0.32\textwidth}
\centering
\includegraphics[width=\linewidth]{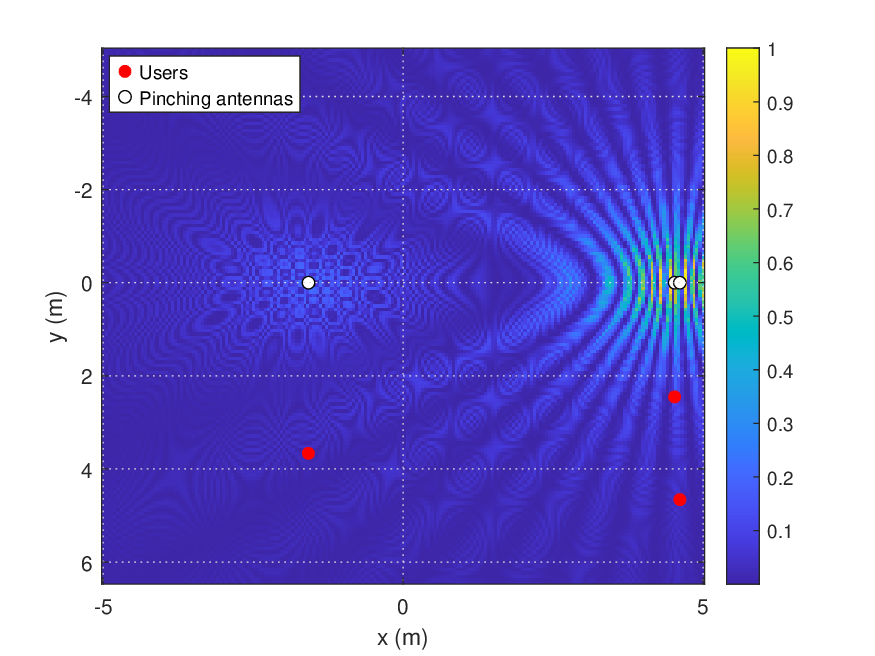}
\vspace{-0.5em}
\caption{NF radiation pattern of the optimized PASS.}
\label{fig:radiation}
\end{minipage}
\end{figure*}

\section{Simulation Results and Discussion}

In this section, we evaluate the performance of the proposed AO-SCA algorithm for a downlink multi-user PASS system. Unless otherwise stated, the BS employs $N=3$ pinching antennas mounted on a Teflon waveguide operating at $28$ GHz and serves $M=3$ single-antenna users. The waveguide length is set to $D=10$ m. The user locations are given by $u_1=(-1.58,-3.66)$, $u_2=(4.523,2.4517)$, and $u_3=(4.6130,4.5873)$, all located within the NF region of the PASS.

Fig.~\ref{fig:geometry} illustrates the optimized pinching antenna positions obtained using Algorithm~\ref{alg:AO_SCA} at $P_t=20\,\text{dBm}$. It can be observed that the antennas are placed in proximity to their associated users, thereby reducing the free-space propagation distances and improving the effective channel gains. This demonstrates the ability of the proposed framework to adapt the antenna placement according to the spatial user distribution and exploit the additional spatial degrees of freedom offered by PASS.

Next, Fig.~\ref{fig:comparison_geo} shows the optimized antenna-user geometries for different system dimensions with $N=M\in\{3,6,9,12\}$ at $P_t=30$ dBm. As the number of antennas and users increases, the spatial deployment becomes denser, enabling the system to exploit additional spatial resources for multi-user transmission. The optimized antenna locations remain closely aligned with the user distribution, highlighting the scalability of the proposed placement strategy under larger network configurations.

Fig.~\ref{fig:objective} depicts the convergence behavior of the proposed AO-SCA algorithm. It can be observed that the achievable sum rate increases monotonically across iterations and converges within a few iterations. This behavior is consistent with the convergence analysis presented in Section~IV, where each AO-SCA iteration yields a non-decreasing objective value. The results confirm the effectiveness of the proposed framework in jointly optimizing the pinching antenna locations and transmit power allocation.

Fig.~\ref{fig:radiation} illustrates the NF radiation pattern generated by the optimized pinching antenna configuration. Owing to the coherent superposition of the electromagnetic fields radiated by the pinching antennas, the radiated power is concentrated around the intended user locations. This focused energy distribution demonstrates the capability of PASS to spatially shape the electromagnetic field and enhance the effective channel gains of the served users while limiting power leakage to unintended directions.

\begin{figure*}[t]
\centering

\begin{minipage}{0.32\textwidth}
\centering
\includegraphics[width=\linewidth]{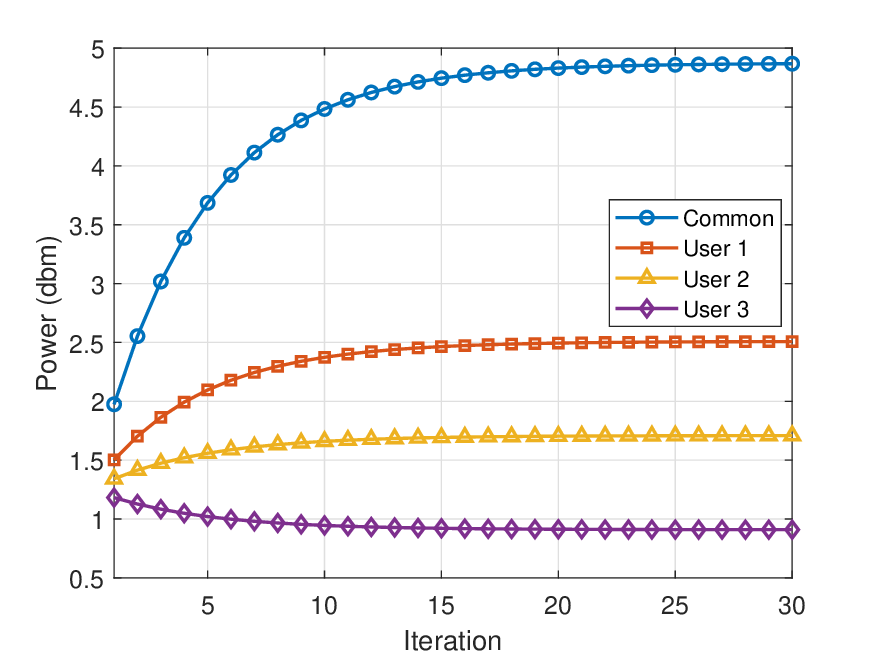}
 \vspace{-0.5em}
\caption{Power allocation convergence.}
\label{fig:power}
\end{minipage}
\hfill
\begin{minipage}{0.32\textwidth}
\centering
\includegraphics[width=\linewidth]{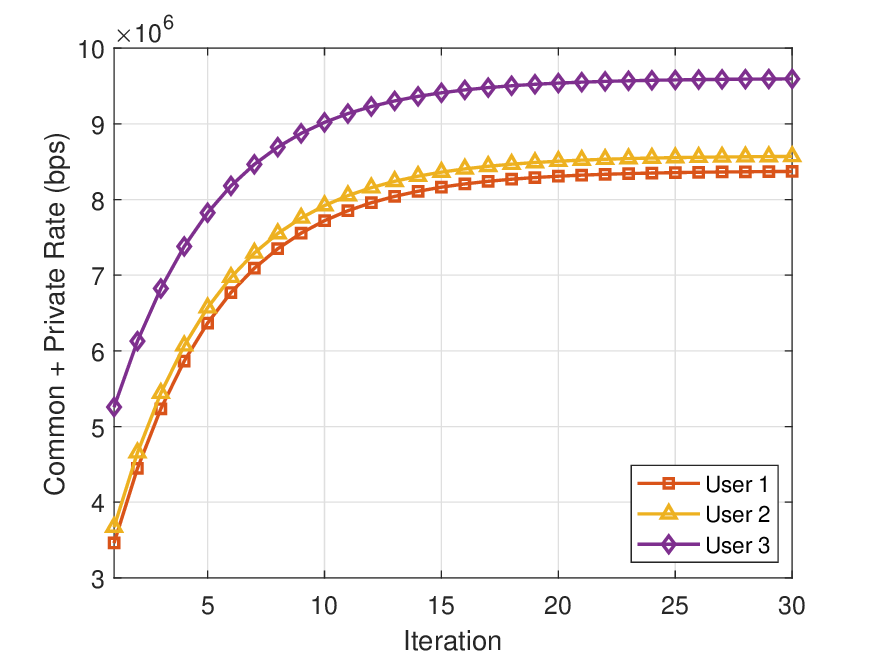}
\vspace{-0.5em}
\caption{Achievable rate convergence.}
\label{fig:common_private_rate}
\end{minipage}
\hfill
\begin{minipage}{0.32\textwidth}
\centering
\includegraphics[width=\linewidth]{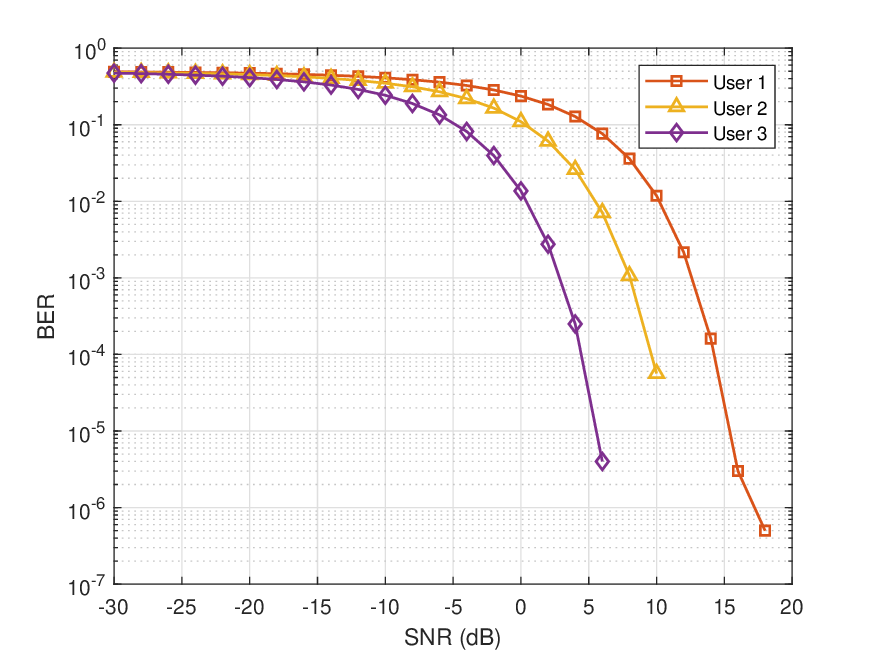}
\vspace{-0.5em}
\caption{BER performance versus SNR.}
\label{fig:ber}
\end{minipage}
\end{figure*}

Fig.~\ref{fig:power} illustrates the evolution of the transmit power allocation during the AO-SCA iterations. It can be observed that the power allocation gradually converges to a stable operating point while satisfying the total power and QoS constraints. A larger fraction of the available transmit power is ultimately assigned to the common stream, reflecting the ability of RSMA to efficiently serve multiple users through a shared transmission layer. By increasing the common-stream power, the system enhances the decodable common rate at all users, thereby improving the overall achievable sum rate while maintaining the required private-stream performance.

Fig.~\ref{fig:common_private_rate} depicts the evolution of the achievable user rates throughout the optimization process. The rates increase monotonically and converge after a few iterations, confirming the effectiveness of the proposed AO-SCA framework. Moreover, the achieved rates remain relatively balanced across users despite their heterogeneous channel conditions. This behavior highlights the interference management capability of RSMA, where the joint transmission of common and private streams enables an effective trade-off between maximizing system throughput and maintaining user fairness.

Fig.~\ref{fig:ber} presents the BER performance of the proposed RSMA-enabled PASS employing quadrature phase-shift keying (QPSK) modulation. As expected, the BER decreases with increasing SNR for all users, demonstrating the reliability gains achieved at higher received signal powers. A performance gap can be observed among users due to differences in their effective channel gains and power allocations. In particular, users with stronger effective channels experience lower BER owing to higher received SINR after common-stream decoding and SIC, whereas users with weaker channels exhibit slightly higher error rates. Nevertheless, all users achieve the characteristic exponential BER decay with SNR, validating the effectiveness of the proposed RSMA-PASS architecture for reliable multi-user transmission.

\section{Conclusion}

We investigated an RSMA-enabled multi-user PASS for downlink communications. A fairness-aware sum-rate maximization problem was formulated to jointly optimize the pinching antenna locations and the power allocation of common and private streams. An AO-SCA framework combining greedy antenna placement and SCA-based power allocation was developed to solve the resulting mixed discrete-continuous non-convex problem. Simulation results demonstrated notable improvements in sum-rate, user fairness, and BER performance compared with conventional non-RSMA PASS schemes. These results highlight the potential of combining PASS and RSMA for future user-centric wireless networks.

\bibliographystyle{IEEEtran}
\bibliography{references}

@article{Mao2018RSMA,
   title={{Rate-splitting multiple access for downlink communication systems: bridging, generalizing, and outperforming SDMA and NOMA}},
   volume={2018},
   ISSN={1687-1499},
    number={1},
   journal={EURASIP J. Wirel. Commun. Netw.},
   publisher={Springer Science and Business Media LLC},
   author={Mao, Yijie and Clerckx, Bruno and Li, Victor O.K.},
   year={2018},
   month={May} }

@ARTICLE{Clerckx2023Primer,
  author={Clerckx, Bruno and Mao, Yijie and Jorswieck, Eduard A. and Yuan, Jinhong and Love, David J. and Erkip, Elza and Niyato, Dusit},
  journal={IEEE J. Sel. Areas Commun.}, 
  title={{A Primer on Rate-Splitting Multiple Access: Tutorial, Myths, and Frequently Asked Questions}}, 
  year={2023},
  volume={41},
  number={5},
  pages={1265-1308},
  doi={10.1109/JSAC.2023.3242718}}

@ARTICLE{Liu2025PASS_Mag,
  author={Liu, Yuanwei and Wang, Zhaolin and Mu, Xidong and Ouyang, Chongjun and Xu, Xiaoxia and Ding, Zhiguo},
  journal={IEEE Commun. Mag.}, 
  title={{Pinching-Antenna Systems: Architecture Designs, Opportunities, and Outlook}}, 
  year={2026},
  volume={64},
  number={1},
  pages={190-196}}

@ARTICLE{Liu2025PASS_Principles,
  author={Yang, Zheng and Wang, Ning and Sun, Yanshi and Ding, Zhiguo and Schober, Robert and Karagiannidis, George K. and Wong, Vincent W.S. and Dobre, Octavia A.},
  journal={IEEE Wireless Commun.}, 
  title={{Pinching Antennas: Principles, Applications and Challenges}}, 
  year={2025},
  volume={},
  number={},
  pages={1-10}}

@ARTICLE{Chen2025PASS_Rate,
  author={Xu, Yanqing and Ding, Zhiguo and Karagiannidis, George K.},
  journal={IEEE Wireless Commun. Lett.}, 
  title={{Rate Maximization for Downlink Pinching-Antenna Systems}}, 
  year={2025},
  volume={14},
  number={5},
  pages={1431-1435}}

@INPROCEEDINGS{Yang2020RSMA_SumRate,
  author={Yang, Zhaohui and Chen, Mingzhe and Saad, Walid and Shikh-Bahaei, Mohammad},
  booktitle={Proc. 2020 IEEE ICC}, 
  title={{Downlink Sum-Rate Maximization for Rate Splitting Multiple Access (RSMA)}}, 
  year={2020},
  volume={},
  number={},
  pages={1-6},
  doi={10.1109/ICC40277.2020.9149417}}

@book{Boyd2004Convex,
  title={Convex Optimization},
  author={Boyd, Stephen and Vandenberghe, Lieven},
  year={2004},
  publisher={Cambridge University Press},
  address={Cambridge, U.K.},
  chapter={11},
  note={Chapter 11: Interior-Point Methods}}

@ARTICLE{10945421,
  author={Ding, Zhiguo and Schober, Robert and Vincent Poor, H.},
  journal={IEEE Trans. Commun.}, 
  title={{Flexible-Antenna Systems: A Pinching-Antenna Perspective}}, 
  year={2025},
  volume={73},
  number={10},
  pages={9236-9253},
 }

@INPROCEEDINGS{9936974,
  author={Jolly, Aditya and Singh, Keshav and Biswas, Sudip},
  booktitle={2021 IEEE ANTS}, 
  title={{RSMA for IRS Aided 6G Communication Systems: Joint Active and Passive Beamforming Design}}, 
  year={2021},
  volume={},
  number={},
  pages={7-12},
  doi={10.1109/ANTS52808.2021.9936974}}

@ARTICLE{8744489,
  author={Singh, Keshav and Wang, Kaidi and Biswas, Sudip and Ding, Zhiguo and Khan, Faheem A. and Ratnarajah, Tharmalingam},
  journal={IEEE Trans. Wirel. Commun}, 
  title={{Resource Optimization in Full Duplex Non-Orthogonal Multiple Access Systems}}, 
  year={2019},
  volume={18},
  number={9},
  pages={4312-4325},
  doi={10.1109/TWC.2019.2923172}}

@ARTICLE{11263923,
  author={Xu, Xiaoxia and Mu, Xidong and Wang, Zhaolin and Liu, Yuanwei and Nallanathan, Arumugam},
  journal={IEEE Trans. on Commun.}, 
  title={{Pinching-Antenna Systems ({PASS}): Power Radiation Model and Optimal Beamforming Design}}, 
  year={2026},
  volume={74},
  number={},
  pages={2160-2175},
  doi={10.1109/TCOMM.2025.3636083}}

\vspace{11pt}
\end{document}